\documentclass[floatfix,aps,prl,showpacs,amsmath,nofootinbib,preprintnumbers,twocolumn]{revtex4}

\usepackage{graphicx}
\usepackage{bm}

\newcommand{\figurewidth}{3.4in}

\def\half{{1\over 2}}

\def\half{{1\over 2}}
\def\({\left(}
\def\){\right)}
\def\[{\left[}
\def\]{\right]}

\def\e{\begin{equation}}
\def\q{\end{equation}}
\def\m{\begin{eqnarray}}
\def\n{\end{eqnarray}}

\begin{document}

\title{Lyth bound revisited}

\author{Qing-Guo Huang}\email{huangqg@itp.ac.cn}
\affiliation{State Key Laboratory of Theoretical Physics, Institute of Theoretical Physics, 
Chinese Academy of Science, Beijing 100190, People's Republic of China}

\date{\today}

\begin{abstract}

Imposing that the excursion distance of inflaton in field space during inflation be less than the Planck scale, we derive an upper bound on the tensor-to-scalar ratio at the CMB scales, i.e. $r_{*,max}$, in the general canonical single-field slow-roll inflation model, in particular the model with non-negligible running of the spectral index $\alpha_s$ and/or the running of running $\beta_s$. 
We find that $r_{*,max}\simeq 7\times 10^{-4}$ for $n_s=0.9645$ without running and running of running, and $r_{*,max}$ is significantly relaxed to the order of ${\cal O}(10^{-2}\sim 10^{-1})$ in the inflation model with $\alpha_s$ and/or $\beta_s\sim +{\cal O}(10^{-2})$ which are marginally preferred by the Planck 2015 data.

\end{abstract}

\pacs{98.80.Cq, 04.30.Db, 98.70.Vc }

\maketitle


Inflation \cite{Starobinsky:1980te,Guth:1980zm,Linde:1981mu,Albrecht:1982wi} is the leading paradigm for the early universe. Not only does it solve several major puzzles of the hot big bang model, such as the flatness puzzle, the horizon puzzle and so on, it offers a natural mechanism to explain the origin of cosmic structure. 
The gravitational waves can be produced in the early universe \cite{Starobinsky:1979ty}, whose direct detection would be taken as the evidence for the inflationary universe and fix the energy scale of inflation.

In March of 2014, BICEP2 detected an excess of B-mode power over the lensed-$\Lambda$CDM expectation on large scales at more than $5\sigma$ CL \cite{Ade:2014xna}. Unfortunately, this excess does not provide evidence for the detection of the primordial gravitational wave once the contribution to the CMB B-mode power from polarized thermal dust was carefully estimated in \cite{Mortonson:2014bja,Flauger:2014qra} and Planck HFI 353 GHz CMB polarization data \cite{Adam:2014bub} were combined with Planck2013 TT and WMAP Polarization data in \cite{Cheng:2014pxa}. Recently the cross-spectra between BICEP2/Keck Array maps and all the polarized bands of Planck (BKP) confirmed that the excess of B-modes detected by BICEP2 mainly originates from the polarized dust \cite{Ade:2015tva}, not the primordial gravitational waves. The combination of the BKP likelihood and Planck 2015 likelihood yields the constraint on the tensor-to-scalar ratio $r$: 
\m
r_{0.002}<0.08
\label{bkp15}
\n 
at $95\%$ CL in \cite{Ade:2015lrj}, where the subscript $0.002$ means that the value of tensor-to-scalar ratio is evaluated at the pivot scale $0.002$ Mpc$^{-1}$. 

As an alternative to the inflationary universe, ekpyrotic scenario proposed by Khoury et al. in \cite{Khoury:2001wf} is also supposed to address the flatness, horizon and monopole puzzles and generate a nearly scale-invariant spectrum of density perturbations. But the ekpyrotic scenario predicts a strongly blue-tilted tensor power spectrum with the spectral index $n_t\equiv d\ln P_t/d\ln k\simeq 2$ which provides an observational signature for distinguishing the ekpyrotic scenario from inflation model which predicts $n_t= -r/8\simeq 0$ \cite{Liddle:1992wi}, where $P_t$ is the amplitude of the tensor power spectrum. 
Even though the current data are not good enough for tightly constraining the tilt of tensor power spectrum, a strongly blue-tilted tensor power spectrum with $n_t\simeq 2$ predicted by the ekpyrotic scenario is disfavored at high confidence level \cite{Huang:2015gka} once the constraint on the intensity of a stochastic gravitational-wave background from the Laser Interferometer Gravitational-Waves Observatory (LIGO) \cite{Aasi:2014zwg} was combined. This provides a strong support to the inflationary universe.

Since the primordial gravitational waves encode critical information about inflationary universe, we wonder what the predictions about the tensor power spectrum in the inflation model are. In \cite{Lyth:1996im}, Lyth proposed that the amplitude of tensor power spectrum is bounded from above by imposing a sub-Planckian excursion of inflaton in field space during inflation. This is the so-called ``Lyth bound". In this letter we will extend the discussion in \cite{Lyth:1996im} to more general cases.

Actually there is a long-term debate about whether the excursion distance of inflaton in field space can go beyond the Planck scale. \\
\noindent $\bullet$ The chaotic inflation model with a monomial potential \cite{Linde:1983gd} is a typical large-field inflation model in which $\phi/M_p\sim {\cal O}(10)$ during inflation. However, according to the weak gravity conjecture for the scalar field proposed in \cite{Huang:2007gk},  the vacuum expectation value of inflaton field should be smaller than the Planck scale, and hence we concluded that such a model is not self-consistent and should be ruled out \cite{Huang:2007gk}. Now it has been disfavoured at around $3\sigma$ CL \cite{Cheng:2014pxa,Ade:2015lrj}. \\
\noindent $\bullet$ Another typical large-field inflation model is the natural inflation model \cite{Freese:1990rb}, where the inflaton is a pseudo-Goldstone boson with a cosine potential. However it requires a super-Planckian decay constant which is not consistent with string theory \cite{Banks:2003sx}. Now this model is also in trouble \cite{Ade:2015lrj}. \\
\noindent $\bullet$ Brane inflation \cite{Dvali:1998pa} provides a generic prototype of inflation in string theory. In \cite{Baumann:2006cd} the authors explicitly derived the microscopic bound on the maximal field variation of inflaton during brane inflation in the warped geometry, and found that it is impossible to achieve a super-Planckian excursion distance of inflaton in field space. \\
In this letter we suppose that the excursion distance of inflaton in field space during inflation be less than the reduced Planck scale $M_p$, and then derive an upper bound on the tensor-to-scalar ratio in the general canonical single-field slow-roll inflation model.

The dynamics of the canonical single-field slow-roll inflation model is govern by 
\m
\ddot \phi+3H\dot \phi+V'(\phi)=0, 
\n 
and 
\m
H^2={1\over 3M_p^2} \(\half \dot\phi^2 +V(\phi)\),
\n
where $V(\phi)$ is the potential of inflaton field $\phi$, and the dot and prime denote the derivatives with respect to the cosmic time $t$ and the inflaton field $\phi$ respectively. The inflaton field slowly rolls down its potential if $\dot\phi^2\ll V$ and $|\ddot \phi|\ll 3H|\dot \phi|$, or equivalently $\epsilon_1\ll 1$ and $|\epsilon_2|\ll 3$, where $\epsilon_1$ and $\epsilon_2$ are the slow-roll parameters defined by 
\m
\epsilon_1&\equiv&{M_p^2\over 2}\({V'\over V}\)^2, \\
\epsilon_2&\equiv&M_p^2{V''\over V}. 
\n
Usually the power spectrum of the scalar perturbation generated by inflation model can be parameterized as follows 
\m
P_s(k)=A_{s,*} \({k\over k_*}\)^{n_{s,*}-1+\half \alpha_{s,*} \ln {k\over k_*}+{1\over 6}\beta_{s,*}({k\over k_*})^2+\cdots},
\n
where $A_{s,*}$, $n_{s,*}$, $\alpha_{s,*}$ and $\beta_{s,*}$ are the amplitude of scalar power spectrum, the spectral index, the running of spectral index and the running of running at the pivot scale $k_*$. In general, the spectral index of scalar power spectrum, the running of spectral index and the running of running are defined by 
\m
n_s&\equiv& 1+{d\ln P_s(k)\over d\ln k}, \label{def:ns}\\
\alpha_s&\equiv& {dn_s\over d\ln k}, \label{def:nrun}\\
\beta_s&\equiv& {d\alpha_s\over d\ln k}. \label{def:nrunn} 
\n
In the canonical single-field slow-roll inflation model, these three parameters are related to the slow-roll parameters by 
\m
n_s&=& 1-6\epsilon_1+2\epsilon_2+2q\epsilon_3+\cdots, \label{ns}\\
\alpha_s&=& -24\epsilon_1^2+16\epsilon_1\epsilon_1-2\epsilon_3-2q\epsilon_4+\cdots, \label{nrun}\\
\beta_s&=& -192\epsilon_1^3+192 \epsilon_1^2\epsilon_2-32 \epsilon_1\epsilon_2^2-24\epsilon_1\epsilon_3 \nonumber \\
&&+2\epsilon_2 \epsilon_3+2\epsilon_4+\cdots,  \label{nrunn}
\n 
up to higher order slow-roll parameters, where $q\simeq 1.063$, and 
\m
\epsilon_3\equiv M_p^4{V'(\phi)V'''(\phi)\over V^2(\phi)}, \quad 
\epsilon_4\equiv M_p^6{V'^2(\phi) V''''(\phi)\over V^3(\phi)}. 
\n
And the tensor-to-scalar ratio is given by 
\m
r\equiv {P_t\over P_s}\simeq 16\epsilon_1 \[1-2q(2\epsilon_1-\epsilon_2)\]. 
\label{ttsr}
\n
See \cite{Stewart:1993bc} and the appendix of \cite{Huang:2014yaa} in detail. Recently Planck collaboration released their scientific results about inflation in \cite{Ade:2015lrj}. The combination of Planck TT, TE, EE and lowP datasets implies 
\m
n_s=0.9645\pm 0.0049, \label{p15ns}
\n
at $68\%$ CL. Adding the running of spectral index, they found 
\m
\alpha_s=-0.0057\pm 0.0071, \label{p15run}
\n
at $68\%$ CL. However, once the running of running is allowed to float, they got 
\m
n_s&=&0.9586\pm 0.0056, \label{p15nsc}\\
\alpha_s&=&0.009\pm 0.010, \label{p15runc}\\
\beta_s&=&0.025\pm 0.013, \label{p15runnc}
\n
at $68\%$ CL. All of the above constraints are evaluated at the pivot scale $k_*=0.05$ Mpc$^{-1}$. Even though the scale-independent spectral index can fit the data, allowing for the running of running provides a better fit to the Planck data, such that $\Delta\chi^2\simeq -4.9$.

In the slow-roll paradigm, the excursion distance of inflaton in field space compared to the reduced Planck scale is related to the tensor-to-scalar ratio by 
\m
{|\Delta \phi |\over M_p}\equiv {|\phi(t_N)-\phi(t_{\rm end})| \over M_p}= {1\over \sqrt{8}} \int_0^N \sqrt{r(N')} dN', 
\label{dphi}
\n
where $t_N$ is the cosmic time corresponding to the number of e-folds $N$ before the end of inflation, which is defined by $N\equiv \int_{t_N}^{t_{\rm end}} Hdt$. 
In general, the tensor-to-scalar ratio $r$ is a function of time (or equivalently, the e-folding number before the end of inflation $N$), and then the bound on $r$ for ${|\Delta \phi |/ M_p}<1$ is model-dependent. In \cite{Huang:2007qz}, the slow-roll parameter $\epsilon_1$ is supposed to be parametrized by 
\m
\epsilon_1(N)=\epsilon_0+{c^2/2 \over N^{2-2\varepsilon}}, 
\label{p1}
\n
where $\epsilon_0$, $c$ and $\varepsilon$ are constants, and $\varepsilon \in [0,1]$. 
Even though this parametrization cannot cover all of the single-field inflation model, it is quite generic. For example, 
for $\epsilon_0=0$, $c^2=p/2$ and $\varepsilon=1/2$, it corresponds to the chaotic inflation model with $V(\phi)\propto \phi^p$; $\epsilon_0=0$, $\varepsilon=1/d$ and $c\ll 1$, it corresponds to the brane inflation model with $V(\phi)=V_0\[ 1-(\mu/\phi)^{d-2}\]$. 
From Eq.~\eqref{p1}, requiring ${|\Delta \phi |/ M_p}<1$ yields $\epsilon_1(N) < 1/(2N^2)$ and then 
\m 
r< {8\over N^2}. 
\label{plb}
\n
The CMB scales roughly corresponds to $N_*\sim 60$ \cite{Liddle:2003as}, and then   
\m
r_*\lesssim 2\times 10^{-3}.
\n
Furthermore, based on this parameterization, we can also get the bounds on the spectral index, the running of spectral index and the running of running, namely  
\m
1-{2\over N}\leq n_s\leq 1,\ -{2\over N^2}\leq \alpha_s\leq 0,\ -{4\over N^3}\leq \beta_s\leq 0. 
\n
See \cite{Huang:2007qz} in detail. 
All of these predictions are nicely consistent with the constraints from Planck 2015, for example those given in Eqs.~\eqref{p15ns} and \eqref{p15run}.

Even though a power-law scalar power spectrum can fit the Planck data well, there is still a big room for allowing a running spectral index, e.g. Eqs.~\eqref{p15nsc}, \eqref{p15runc} and \eqref{p15runnc}. In order to include these results, we need to go beyond the parametrization in Eq.~\eqref{p1}. 
For simplicity, we introduce a new variable defined by 
\m
x=\ln {a\over a(t_*)},
\n
where $t_*$ denotes the time of horizon exit of the perturbation mode $k_*$ during inflation. 
Since $r\equiv P_t/P_s$, 
\m
{d\ln r \over dx}=n_t-(n_s-1). 
\n 
Considering the consistency relation $n_t=-r/8$, we have 
\m
{d\ln r\over dx}=-\[(n_s-1)+{r\over 8}\]. 
\label{drx}
\n 
For $n_{s,*}=0.9645$ and $r_*<0.08$, $r$ increases with the expansion of the universe. However, if there are non-negligibly positive running of the spectral index and/or running of running, $n_s$ increases  and the right hand side of the above equation may flip the sign after a few number of e-folds, and then $r$ drops down. Therefore one may expect that the Lyth bound is expected to be significantly relaxed in the case with non-negligibly positive running and/or running or running. 
Since the perturbation mode $k$ exits horizon when $k=aH$ and $H$ is roughly a constant during inflation, $d\ln k\simeq d\ln a=dx$. 
Integrating over Eqs.~\eqref{def:nrun} and \eqref{def:nrunn}, we obtain 
\m
\alpha_s&=&\alpha_{s,*}+\beta_{s,*} x, \\
n_s&=&n_{s,*}+\alpha_{s,*}x+\half \beta_{s,*} x^2. \label{nsab}
\n
Therefore 
\m
{d\ln r \over dx}\simeq -\[(n_{s,*}-1)+{r_*\over 8}+\alpha_{s,*}x+{\beta_{s,*}\over 2}x^2 \] . 
\n 
Here $r$ is replaced by $r_*$ on the right hand side of the above equation because $|d\ln r/dx| \ll 1$ and $r=r_*+r_* \(d\ln r/dx\)|_{r=r_*} x+\cdots\simeq r_*$. 
Integrating over the above equation, we obtain 
\m
r\simeq r_* e^{-R(n_{s,*},r_*,\alpha_{s,*},\beta_{s,*};x)}, 
\n
where 
\m
&&R(n_{s,*},r_*,\alpha_{s,*},\beta_{s,*};x)  \\
&&\simeq \[(n_{s,*}-1)+{r_*\over 8}\]x +\half \alpha_{s,*} x^2+{\beta_{s,*}\over 6}x^3. \nonumber
\n
Now Eq.~\eqref{dphi} reads 
\m
{|\Delta \phi|\over M_p}=\sqrt{r_*\over 8} \int_0^{N_*} dx \exp \[ -\half R(n_{s,*},r_*,\alpha_{s,*},\beta_{s,*};x) \]. \nonumber \\ 
\n
Requiring that the right hand side of the above equation be less than one yields the Lyth bound for the general canonical single-field slow-roll inflation model. Or equivalently, the upper bound on the tensor-to-scalar ratio $r_{*,max}$ satisfies 
\m
r_{*,max}={8\over \( \int_0^{N_*} dx \exp \[ -\half R(n_{s,*},r_{*,max},\alpha_{s,*},\beta_{s,*};x) \] \)^2} . \nonumber \\ 
\label{rmax}
\n
This is the key result of this letter. 
According to Eq.~\eqref{nsab}, one may worry about that the spectral index significantly deviates from unity when $x\gg 1$ if the running of spectral index and/or running of running are large, and then the slow-roll approximation adopted in the previous calculations is not reliable any more.  However, if the running of spectral index and/or running of running are positive and large, the tensor-to-scalar ratio decreases rapidly and becomes quite small after a few e-folding numbers stating from the time of $t_*$. In this case the integration in \eqref{rmax} mainly comes from integrating over these few e-folding numbers during which the slow-roll approximation is still applicable. So our formula \eqref{rmax} is still reliable for the case with non-negligibly positive running of spectral index and running of running.

As a double check, we also adopt the Monte-Carlo simulation to figure out the Lyth bound. Selecting the initial slow-roll parameters at the pivot scale which satisfy the constraints from observations and numerically solving the flow equations of the slow-roll parameters, i.e. 
\m
{d\epsilon_1\over dx}&\simeq& 2\epsilon_1(2\epsilon_1-\epsilon_2),\\
{d\epsilon_\ell \over dx}&\simeq& 2(\ell-1)\epsilon_1 \epsilon_\ell-(\ell-2)\epsilon_2\epsilon_\ell-\epsilon_{\ell+1}, 
\n
where $\ell\geq 2$, we can generate a vast collection of inflation models in which $|\Delta \phi|/M_p<1$ and $\epsilon_1(x)<1$ for $x\in [0,N_*]$. As long as the number of models in such a collection is large enough, we can figure out the upper bound on the tensor-to-scalar ratio. We will compare the Monte-Carlo simulation to the analytic formula \eqref{rmax} in the following three cases respectively.

\begin{figure}[hts]
\begin{center}
\includegraphics[width=\figurewidth]{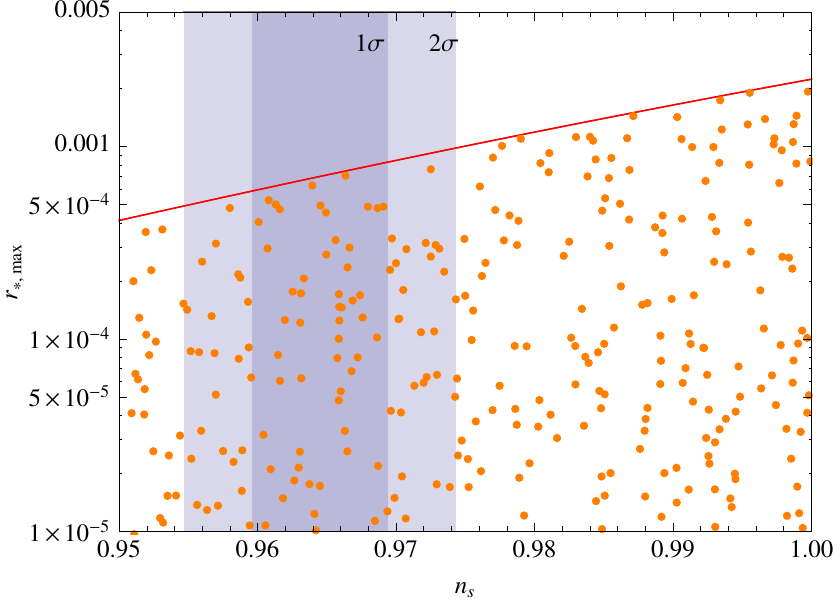}
\end{center}
\caption{The upper bound on the tensor-to-scalar ratio $r_{*,max}$ is illustrated by the red curve. The grey and light grey bands respectively correspond to the constraints on $n_{s,*}$ at $1\sigma$ and $2\sigma$ CL from Planck in Eq.~\eqref{p15ns}. The orange points denote the predictions of a vast collection of models generated by the Monte-Carlo simulation, in which $|\Delta \phi|/M_p<1$ and $\epsilon_1(x)<1$ for $x\in [0,N_*]$. 
}
\label{fig:rns}
\end{figure}

First of all, let's consider the case with negligible $\alpha_{s,*}$ and $\beta_{s,*}$, i.e. $\alpha_{s,*}\simeq \beta_{s,*}\simeq 0$. If $n_{s,*}\neq 1$, Eq.~\eqref{rmax} reads 
\m
r_{*,max}\simeq 2(1-n_{s,*})^2 \[ e^{(1-n_{s,*})N_*/2}-1\]^{-2}. 
\label{rns}
\n 
In the limit of $n_s\rightarrow 1$, $r_{*,max}\simeq 8/N_*^2$ which is same as that in Eq.~\eqref{plb}. 
For $N_*=60$ and $n_{s,*}=0.9645$, we have $r_*<r_{*,max}\simeq 7\times 10^{-4}$. 
\footnote{For $n_{s,*}=0.96$, $r_{*,max}\simeq 6\times 10^{-4}$. Similar to \cite{Huang:2007qz}, the authors considered a specific assumption of $\epsilon_1(N)$ scaling as a power of $1/N$, and found a more stringent bound on the tensor-to-scalar ratio, i.e. $r\lesssim 2\times 10^{-5}$ for $n_s=0.96$, in \cite{Garcia-Bellido:2014wfa}. However, in this paper,  we do not assume any specific parameterization, and hence our results are model-independent. 
}
The upper bound on the tensor-to-scalar ratio for the case with $n_{s,*}<1$ is smaller than that for the Harrison-Zel'dovich spectrum $(n_{s,*}=1)$. 
The upper bound on the tensor-to-scalar ratio is illustrated by the red curve in Fig.~\ref{fig:rns}.  
We can also check Eq.~\eqref{rns} by using the Monte-Carlo simulation. 
Since $\alpha_s\simeq \beta_s\simeq 0$ in this case, the slow-roll parameters can be truncated as $\epsilon_\ell\simeq 0$ for $\ell\geq 3$. Following the method described in the former paragraph and imposing $|\Delta \phi|/M_p<1$ and $\epsilon_1(x)<1$ for $x\in [0,N_*]$, we generate a vast collection of models which are denoted by the orange points in Fig.~\ref{fig:rns}. From Fig.~\ref{fig:rns}, we see that all of the predictions of these models stay below the bound given in Eq.~\eqref{rns}. It implies that the formula in Eq.~\eqref{rns} is reliable.

\begin{figure}[hts]
\begin{center}
\includegraphics[width=\figurewidth]{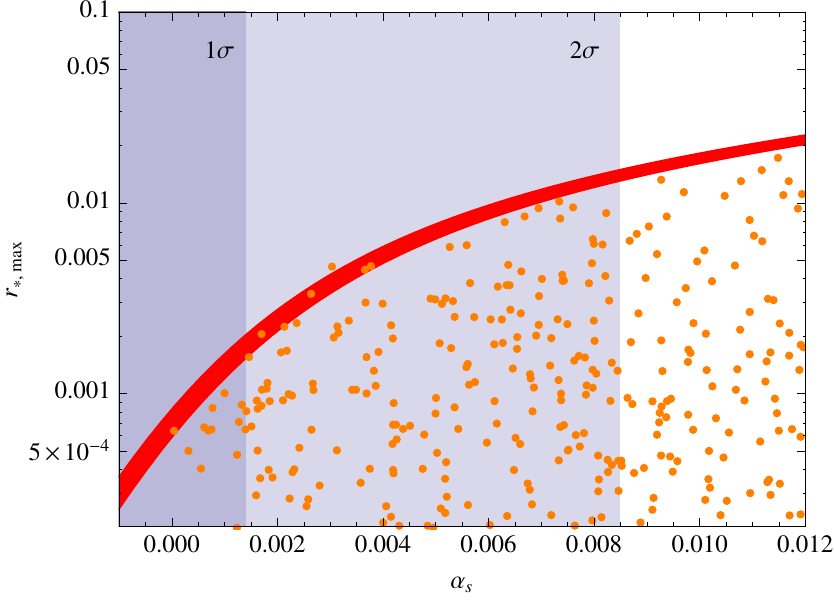}
\end{center}
\caption{The upper bound on the tensor-to-scalar ratio $r_{*,max}$ in the model with non-negligible running of spectral index. The red-shaded region comes from the uncertainty of $n_{s,*}$ in Eq.~\eqref{p15ns}. The grey and light grey bands respectively correspond to the constraints on $\alpha_{s,*}$ at $1\sigma$ and $2\sigma$ CL from Planck in Eq.~\eqref{p15run}. The orange points denote the predictions of a vast collection of models generated by the Monte-Carlo simulation, in which $|\Delta \phi|/M_p<1$ and $\epsilon_1(x)<1$ for $x\in [0,N_*]$. 
}
\label{fig:ralpha}
\end{figure}

Secondly, we consider the model with non-negligible running of spectral index, but $\beta_{s,*}\simeq 0$. Taking $N_*=60$, we numerically solve Eq.~\eqref{rmax} and the results show up in Fig.~\ref{fig:ralpha} where the red-shaded region comes from the uncertainty of $n_{s,*}$ in Eq.~\eqref{p15ns}. 
The upper bound on the tensor-to-scalar ratio is relaxed to be $r_{*,max}\simeq 2\times 10^{-3},\ 1.5\times 10^{-2}$ once the 1 and 2 $\sigma$ uncertainties of the running of spectral index given in Eq.~\eqref{p15run} are taken into account. 
Similar to the previous case, the slow-roll parameters are truncated as $\epsilon_\ell\simeq 0$ for $\ell\geq 4$ and a vast collection of models denoted by the orange points in Fig.~\ref{fig:ralpha} are generated by the Monte-Carlo simulation. Since almost all of the orange points do not exceed the red-shaded region, the analytic formula \eqref{rmax} is reliable for the model with non-negligible running of spectral index. 

\begin{figure}[hts]
\begin{center}
\includegraphics[width=\figurewidth]{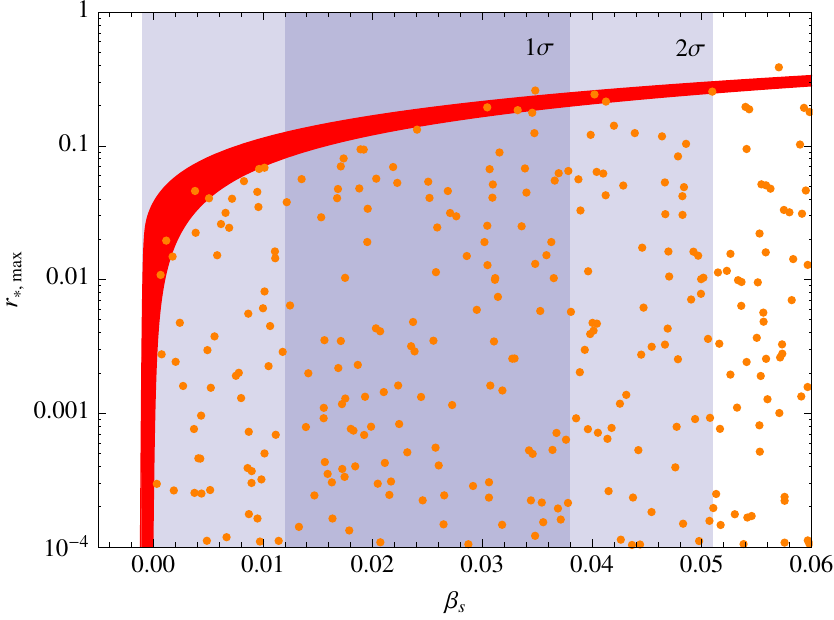}
\end{center}
\caption{The upper bound on the tensor-to-scalar ratio $r_{*,max}$ in the model with non-negligible running of spectral index and running of running. The red-shaded region comes from the uncertainty of $n_{s,*}$ and $\alpha_{s,*}$ in Eqs.~\eqref{p15nsc} and \eqref{p15runc}. The grey and light grey bands respectively correspond to the constraints on $\beta_s$ at $1\sigma$ and $2\sigma$ CL from Planck in Eq.~\eqref{p15runnc}. The orange points denote the predictions of a vast collection of models generated by the Monte-Carlo simulation, in which $|\Delta \phi|/M_p<1$ and $\epsilon_1(x)<1$ for $x\in [0,N_*]$. }
\label{fig:rbeta}
\end{figure}

Third, we consider the case with both non-negligible $\alpha_{s,*}$ and $\beta_{s,*}$. Our numerical results are illustrated in Fig.~\ref{fig:rbeta} where the red-shaded region comes from the uncertainty of $n_{s,*}$ and $\alpha_{s,*}$ in Eqs.~\eqref{p15nsc} and \eqref{p15runc}.
Taking the $1\sigma$ uncertainty of $\beta_{s,*}$ in Eq.~\eqref{p15runnc} into account, we find $r_{*,max}=0.25$ which says that there is still a big room for detecting the primordial gravitational waves generated during inflation if the running of running is in the order of $+{\cal O}(10^{-2})$ which is marginally preferred by Planck 2015 \cite{Ade:2015lrj}. Again, the Monte-Carlo simulation $(\epsilon_\ell\simeq 0,\ \hbox{for}\ \ell\geq 5)$ confirms the reliability of our analytic formula \eqref{rmax} in this case.

To summarize, we derive the upper bound on the tensor-to-scalar ratio for the general canonical single-field slow-roll inflation model with sub-Planckian excursion distance of inflaton in field space. Comparing with the Monte-Carlo simulation, we see that the analytic formula \eqref{rmax} works quite well for the case with non-negligible running of spectral index and/or running of running. Without adding the running of running, a power-law scalar power spectrum give a good fit to the Planck 2015 data. For $N_*=60$ and $n_{s,*}=0.9645$, the tensor-to-scalar ratio is bounded from above by $7\times 10^{-4}$ which might be marginally detected in the future. Adding the running of running provides a better fit to the Planck data, the upper bound is relaxed to be order of ${\cal O}(0.1)$, and there is still a big room for detecting the primordial gravitational waves from both the ground-based experiments and satellites in the near future \cite{Creminelli:2015oda}. See more related discussion about the bound on the tensor-to-scalar ratio for the inflation model with sub-Planckian excursion in \cite{lr}.


\noindent {\bf Acknowledgments}

This work is supported by the project of Knowledge Innovation Program of Chinese Academy of Science and grants from NSFC (grant NO. 11322545 and 11335012).



\end{document}